# Social Media Mining Toolkit (SMMT)

**Ramya Tekumalla and Juan M. Banda***
**Georgia State University**
**25 Park Place, Atlanta, Georgia, 30303**
**Atlanta, Georgia, USA**

**ORCID: Ramya Tekumalla - 0000-0002-1606-4856**
**Juan M. Banda - 0000-0001-8499-824X**

**Running Title: Social Media Mining Toolkit (SMMT)**

**\* Contact: jbanda@gsu.edu**
**Office: +1-404-413-5585**
**FaxL +1-404-413-5717**
**Website: www.jmbanda.com**

**Abstract**

There has been a dramatic increase in the popularity of utilizing social media data for research purposes within the biomedical community. In PubMed alone, there have been nearly 2,500 publication entries since 2014 that deal with analyzing social media data from Twitter and Reddit. However, the vast majority of those works do not share their code or data for replicating their studies. With minimal exceptions, the few that do, place the burden on the researcher to figure out how to fetch the data, how to best format their data, and how to create automatic and manual annotations on the acquired data. In order to address this pressing issue, we introduce the Social Media Mining Toolkit (SMMT), a suite of tools aimed to encapsulate the cumbersome details of acquiring, preprocessing, annotating and standardizing social media data. The purpose of our toolkit is for researchers to focus on answering research questions, and not the technical aspects of using social media data. By using a standard toolkit, researchers will be able to acquire, use, and release data in a consistent way that is transparent for everybody using the toolkit, hence, simplifying research reproducibility and accessibility in the social media domain.

**Keywords:** Data Mining, Machine Learning, Social Media, Information Storage and Retrieval

## Introduction

Only in the last six years, there has been a great influx of research works that describe different types of research works using Twitter and Reddit data, nearly 2,500 papers are found in PubMed [1]. These works encompass countless applications, such as the usage of opioids [2], the flu [3], eating disorder [4] networks analyses, depression symptoms detection [5], and diabetes interventions [6] etc. While all the listed studies use data from Twitter and Reddit, we can only find code available for one of them. Additionally, the data acquisition methodology is different on each study and seldomly reported, a crucial step towards reproducibility of any of their analyses. When it comes to using Twitter data for drug identification and pharmacovigilance tasks, authors of works like [7–9] have been consistently releasing publicly available datasets, software tools, and complete Natural Language Processing (NLP) systems with their works. In an attempt to shift the biomedical community into better practices for research transparency and reproducibility, we introduce the Social Media Mining Toolkit (SMMT), a suite of tools aimed to encapsulate the cumbersome details of acquiring, preprocessing, annotating, and standardizing social media data. The need for a toolkit like SMMT arose from our work using Twitter data for the characterization of disease transmission during natural disasters [10] and mining large-scale repositories for drug usage related tweets for pharmacovigilance purposes [11]. We originally wanted to use other researcher's tools and surprisingly we found very little code available with the majority outdated and non-functioning. Going one step back, we did find rudimentary Python libraries to interact with the Twitter API, but some of their learning curves are steep and not overly documented. We then decided to clean and integrate our code into a toolkit that would help us provide a comprehensive resource for other researchers/users to replicate our work and to use in their own analyses, hence SMMT was born. Parallel to tools like SMMT, there are other research groups that are outlining frameworks to streamline the mining of social media like Sarker et. al. [12], which are complementary to the use and need of this tool.

**Methods**

Programmed using Python version 3 and the latest Twitter API interfaces, the functionality of SMMT is divided into three separate sets of tools: data acquisition tools, data preprocessing tools, and data annotation and standardization tools. The particular versions of the additional Python libraries used by SMMT are available at the github documentation [13] since they are constantly updated and refreshed. Besides extensive usage documentation, the tool also provides two end-to-end usage examples, as well as additional Google Colaboratory [14] interactive Python notebooks with data usage examples. Note than in order to use most of the functionality of SMMT, users need to sign-up to acquire access to the Twitter Application Program Interface API. Once approved, users will be provided a set of API credential keys, more information can be found in [15]. Figure 1 shows all current components of SMMT. In the following sections we provide additional details of each category of available tools.

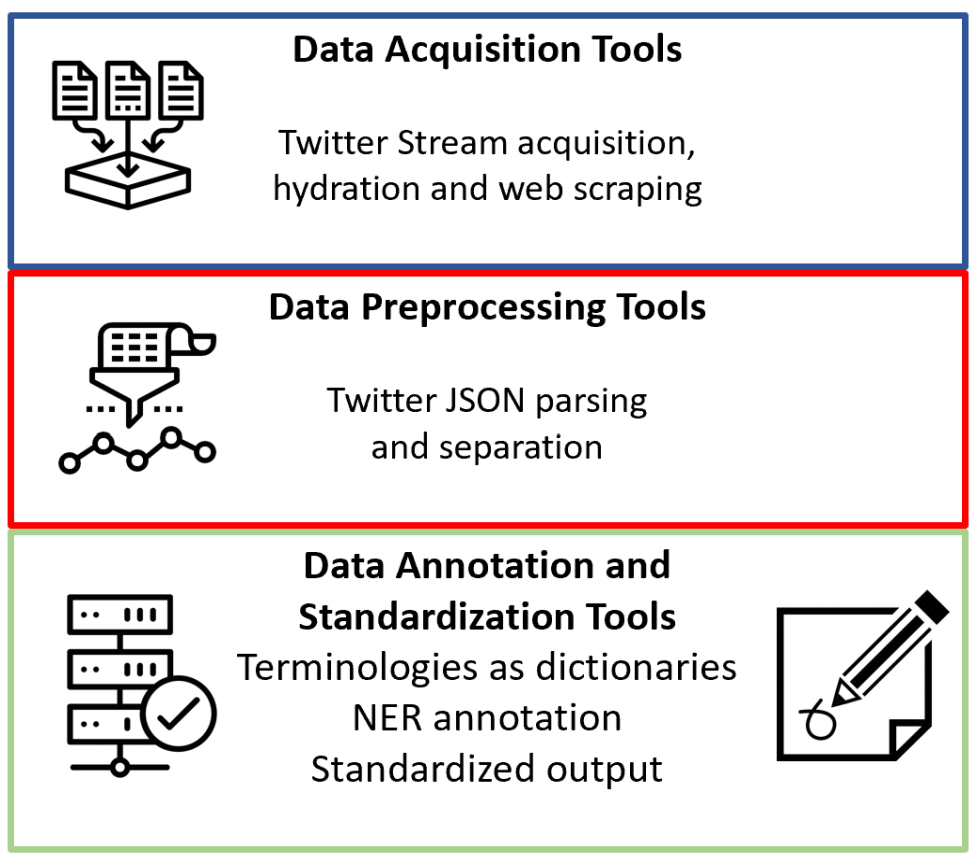

Figure 1. SMMT Tools categories and purpose outline.

**Data Acquisition Tools**

The tools in this category are used to gather data from social media sites, namely Twitter for this initial release of SMMT. The most common way of acquiring Tweets is to use the Twitter streaming API [16]. Our toolkit provides two separate utilities to capture streaming data, one will gather all available tweets and will continue running until terminated (*streaming.py*), and the other will take a list of search keywords and number of desired tweets and will pull those from the current tweet stream (*search_generic.py*). Details on how to use these utilities can be found on the READ.ME file.

The most common and permitted way of sharing Twitter data publicly is by only sharing the tweet id number. This number then needs to be 'hydrated', which means that the Twitter API needs to be used to fetch all the complete tweet and additional meta-data fields. This is a vital step for most users trying to replicate other studies or analyses. We provide a utility called *get_metadata.py* which reads a list of tweet ids and hydrates them automatically.

One of the major drawbacks of the Twitter API is the fact that unless having paid access to it, researchers cannot extract all historical tweets for any given Twitter user. Also, extracting all tweets from a given time range is not always easily and efficiently possible with the API. For these purposes we provide a utility called *scrape.py* which, once given a list of Twitter handles and corresponding date ranges, will automatically scrape the Twitter page and pull the tweet ids of the desired user and date range. These tweet ids then need to be 'hydrated' to be able to fully use them.

**Data Preprocessing Tools**

After having acquired enough data for research purposes from the Twitter stream, or identified and 'hydrated' a publicly available dataset, there is a need to subset the tweets and process the tweets JSON files to extract the fields of interest. While seemingly trivial, most biomedical researchers do not want to work with JSON objects, and since around 70% of the JSON fields are not populated, precise preprocessing steps need to be carried out to clean the data and render it useful in friendlier formats.

SMMT contains the *parse_json_lite.py* tool which takes a relatively small file (less than 1 Gigabyte in size) of Twitter JSON objects and separates these objects into a tab delimited file with each JSON field converted to a column and each tweet into a data row. With over 170 fields of meta-data, researchers are usually not interested in the vast majority of them. This tool can be configured to select which fields are of interest to be parsed and only process those into the tab delimited format. If the size of the tweets JSON objects file is larger than 1 Gigabyte, we provide an additional tool, *parse_json_heavy.py*, which can handle Terabyte sized files sequentially rather than reading them all in memory for speed.

Once all the tweets are processed into the cleaner tab delimited format, which can even be read in Excel, there might be a need to further subset the tweets based on a given list of terms, or dictionary. For this purpose, we have included the *separate_tweet_tsv.py* file, which takes a term list in a format specified in the READ.ME file of SMMT and will return only the tweets that contain the provided terms.

**Data Annotation and Standardization Tools**

After preprocessing the acquired social media data, researchers have the capabilities of standardizing their tweets' text with our set of tools. Taking advantage of OntoGene Bio Term Hub [17] and their harmonization of biomedical terminologies into a unified format, we provide a tool, *create_dictionary.py*, that converts their downloads into SMMT-compatible dictionaries. To avoid creating a complicated and cumbersome format for our tool, we opted for simplicity and only rely on having a tab delimited file with an identifier column and a term name column. Other dictionaries that we have made available will standardize any annotations using the Observational

Health Data Sciences and Informatics (OHDSI) vocabulary [18]. We are testing functionality to also convert our dictionaries to the PubDictionaries [19] format for the next release, allowing researchers to use their functionality and online REST services.

One of the most important tools of SMMT is the Spacy [20] NER annotator, *SMMT_NER_basic.py*, this tool will take the tab delimited tweets, a dictionary file, and the name of the output file for the annotations. In order to extend the usability of our tool, we provide the resulting annotations in a traditional format: document, span, term format; as well as pre-formatted outputs compatible with the brat annotation tool [21] and PubAnnotation and its viewer TextAE [22] as shown in Figure 2.

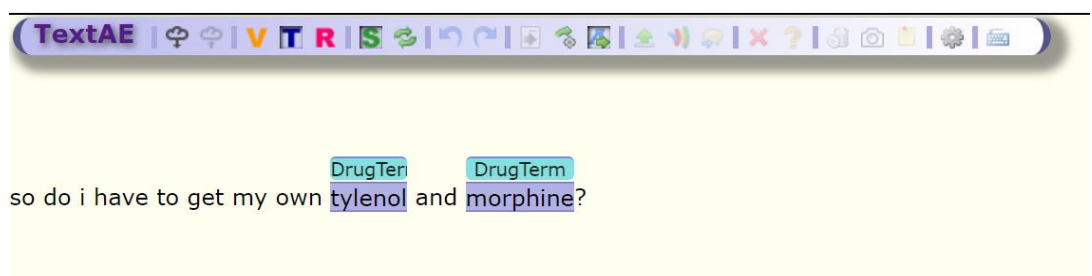

Figure 2. Sample SMMT_NER_basic annotation using an RxNorm-based dictionary and displayed on TextAE.

**Discussion**

While all the tools have their own documentation, in order to ease the adoption of the tools available in SMMT, we have included an end-to-end example in the examples folder that performs the following tasks:

1) Download 300 tweets from the Twitter API stream for each of the following keywords: donald trump,coronavirus,cricket

2) We then preprocess those tweets to extract Tweet Id and their text into tab delimited files.

3) Using a Google Colab Notebook, we use these preprocessed files and then use the TF-IDF vectorizer on the text of the tweets to create a test and train set and build a Multi Nomial Naive Bayes Classifier to separate tweets based on their label. All details and steps of this process are outlined in the Colab Notebook.

4) We then test our trained model on the test set and generate a confusion matrix heat-map (Figure 3) of the classification task, and show the model performance metrics.

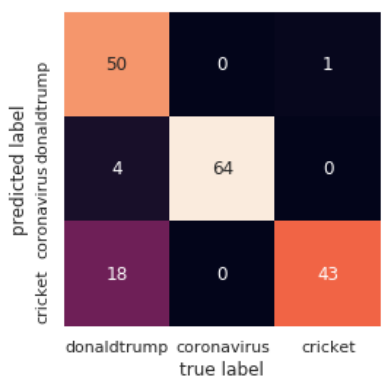

Figure 3. Confusion Matrix Heat-Map for the classification of tweet labels.

The whole process of this example takes less than 30 minutes to complete and is heavily documented for SMMT users to overcome the learning curve of acquiring and preprocessing tweets. While our example is simple in nature, users can build upon it and modify it to better suit their needs.

The tools part of SMMT allow users to simplify their research workflows and to focus on determining which data they want to use and the analyses they want to perform, rather than deciphering how to acquire the data. While most cutting-edge and near real-time research will be done pulling tweets from the Twitter API stream, there are countless datasets available for historical research, from large general purpose databases like the Internet Archive's Twitter Stream Grab dataset [23], which consists of data from 2014 to 2019, to more specialized and pre-curated datasets for uses like Pharmacovigilance [11] among others. This initial version release of SMMT will continue growing with additional tools being developed for platforms like Reddit, Dark Web forums, and other social media data sources.

## Conflicts of Interest

No potential conflict of interest relevant to this article was reported.

**Authors' contribution:**

Conceptualization, Data curation, Formal analysis, Methodology, Writing – original draft, Writing – review & editing: RT and JMB

**Acknowledgments**

We would like to acknowledge Jin-Dong Kim, DBCLS, and ROIS for making our participation in BLAH6 (Biomedical Linked Annotation Hackathon) possible. We also like to thank Javad Rafiei Asl, Kevin Bretonnel Cohen, Núria Queralt Rosinach, Yue Wang and Atsuko Yamaguchi for the help and input during the Biomedical Linked Annotation Hackathon 6.